\documentclass{article}

\newcommand{\be}{\begin{equation}}
\newcommand{\ee}{\end{equation}}
\begin{document}

\title{Lattice Field Theory: past, present and future.}

\author{H. Neuberger\\
Department of Physics and Astronomy\\ Rutgers University,
Piscataway, NJ 08855}
\maketitle

\begin{abstract}
This letter is in response to a recent review by DeTar and Gottlieb
about lattice QCD that has recently appeared in Physics Today. 
It also is partially motivated by a separate review written by
DeGrand. My basic point is that one should be more responsible
when presenting numerical results as coming from {\sl ab initio}
calculations. In turn, this leads me to the suggestion that
in lattice field theory one should go back to computational
support going directly to small groups, 
including groups containing only a single
researcher, rather than concentrating most of the funds into
the hands of few large collaborations.
\end{abstract}

DeTar and Gottlieb's article~\cite{phystod} misleads 
readers from outside lattice field theory about its past, present and 
future. At best, it may present a consensus in 
the collaboration the authors are part of, known as MILC. 

The terms ``lattice field theory (LFT)'', ``lattice gauge theory (LGT)'' and
``lattice QCD (LQCD)'' are often used interchangeably. I shall follow this
practice; the more accurate term will be clear from the context. Also,
I have not put in references to any papers except the ones I quote from
{\sl verbatim}.

After an overview of the history of LFT the authors of~\cite{phystod} 
state that ``the most important theoretical advance in recent years''
was that of ``improvement''. I disagree with this qualification.
In their historical overview the authors miss too many of the field's
truly remarkable achievements. The description below briefly mentions
{\it some} of these achievements and is intended to show 
that ``improvement''  does not measure up. 

The renormalization group (RG),
a conceptual organizing principle of all field theories, 
has been first concretely 
formulated in LFT. Nowadays we have beautiful examples of intricate flow
patterns between different fixed points of continuum field theories showing 
that these concepts actually work to full extent. As a concrete example of the
value of numerical LFT let me mention one non-gauge result of relevance to 
particle physics: triviality is 
indeed a property of scalar field theories and implies a 
non-perturbative bound
on the Higgs mass of the order of $700~ GeV$. Another conceptual 
idea which started its
life in LFT is the concept of duality. Two dual field theories typically 
have different fields, different symmetries, but, if the coupling in
one theory is set to the inverse coupling in the second theory they become
equivalent descriptions of the same physical entity. This is again a concept 
that has been
dramatically validated by marvelous modern examples. 
In the context of duality, 
monopoles have played a central role: again something LFT 
can be proud of (including
some of today's members of MILC). 
Moving on, we recall the conceptual pictures of 
confinement and finite temperature
deconfinement. The first decade of LFT has been 
extremely productive and we
see that it has had a long lasting impact on 
theoretical particle physics and field theory.
Abstract concepts have successfully been buttressed 
by concrete calculations and
accurate and reliable numbers were produced by 
simulations of bosonic degrees of
freedom. One could call this the ``bosonic era of LFT'' 
and it is an illustrious one. 

The inclusion of fermions, a much needed step beyond the bosonic era, 
has preoccupied a large fraction of the community, in various forms.
Fermions had a conceptual defect in their original
formulation by Wilson. A different and ingenious formulation, 
by Kogut and Susskind, 
had defects of a similar kind. For twenty years it was believed 
that continuum flavored
chiral symmetries could not be reproduced on the lattice, 
although they seemed to have 
nothing fundamental to do with continuum ultraviolet physics. 
(Chiral symmetries are important
in particle physics because they provide a key mechanism for 
naturally small masses.)
During the last decade, finally, the problem of lattice chirality 
has been solved. A large number of papers and their citation
counts attest to the interest this development was received with. 
Thus, without making claims
about how this measures up to the bosonic era, this was an 
important theoretical advance in
recent years. 

On the numerical front, progress on fermion systems was 
held in check by computer 
technology. In the US, a very major fraction of computer 
resources was given to few 
large collaborations, and MILC is the oldest among these. 
Progress was made at essentially the rate that computer power
got cheaper. On the way, a significant physical step was the 
formulation of the
valence approximation and establishing its surprising numerical 
accuracy when compared
to experiment. At the more technical level, probably 
the most significant step was the
discovery of an algorithm that would be able to take us 
beyond the valence approximation,
to truly {\sl ab initio} numerical QCD. Neither of these 
two steps originated from MILC,
although MILC made contributions at later stages. 
Both the surprises surrounding
the valence approximation, and the algorithms making 
it feasible to go beyond it,
would qualify as important advances, whether ``theoretical'' or not is
a matter of semantics.

Since the initial formulation of LFT, and the associated 
RG ideas, it was clear that
the approach to continuum could be sped up by 
fine-tuning the lattice action. This is
the improvement technique, and it is very important in practice, but it
does not have the theoretical novelty any of the above achievements have.

The newest results the article describes go
beyond the valence approximation and use clever improvement tricks.
As a result, if we ignore one less appealing trick that is being used,
we could say that improvement has made the corrections to the continuum
limit vanish as $\frac{a^2}{|\log [ a \Lambda_{\rm QCD} ]|}$ 
rather than $a^2$, where $a$ is the
lattice spacing of the grid. The improvement trick is 
certainly valid research, but
not a breakthrough and certainly not the ``most important 
theoretical advance''  
of recent years.

However, the real problem, and one that the article hides 
from the unsuspecting reader, 
is that these recent calculations are not {\sl ab initio}
because of ``the less appealing trick'' used to include 
sea quarks. It is true that ordinary effective field theory (EFT) 
logic would support the use of
simulations based on Kogut-Susskind fermions 
for gauge theories with four-fold
degenerate flavors. In other words, if each flavor were 
to come in four identical ``tastes'',
one could make a case, at least on the level of 
EFT logic, that,
so long as asymptotic freedom is preserved, we 
can, in principle, carry out an {\sl ab initio}
numerical calculation for observables in this 
field theory. However, the simulations are
claimed to describe QCD, and there is no such 
degeneracy in QCD. For this reason, the simulations described
employ a trick in which the contributions of 
sea quarks are reduced by a factor of four.
However, even a generous application of 
EFT lore, defending this
procedure, has never been proposed. If we did not carry out this artificial
suppression by one quarter, the effective lagrangian would have
terms that violate taste equivalence at 
order $\frac{a^2}{|\log [ a\Lambda_{\rm QCD} ]|}$. There is no
place I know of that sketches a derivation of an EFT 
description for what is actually being simulated, 
inclusive of the artificial factor of
one quarter suppression of sea quark contributions. 

Moreover, EFT logic is usually applied
by looking only at the local properties of the theory and can
go wrong if there is an important global property that has
not been properly taken into account. In our case there would
be reason to be cautious even if an EFT 
argument supporting the artificial one quarter suppression
were presented: Apparently, Kogut - Susskind fermions can be
viewed as Grassmann valued antisymmetric forms and, as such,
could be defined on a lattice approximating a manifold, like
$CP^2$, which does not admit a spin structure and, therefore,
cannot accommodate four non-interacting (in a fixed background)
degenerate Dirac fermions. It is unlikely that the fermion
determinant has an acceptable fourth root in this case. Typically,
EFT is applied without paying attention to global topology, so,
at least in this example, could easily have led us astray.

Contrary to the impression the article~\cite{phystod} conveys, 
a right way to full QCD is known,
based on exact lattice chirality. Again, 
computational cost holds us back, but
there is little doubt that more years will 
bring us the power we need to
do the calculations right. It seems now, and, 
if one is so inclined, 
the recent result reviewed in the article~\cite{phystod}
might be taken as supportive of this hypothesis, 
that during the next ten years,
or so, we shall be able to present to the rest of the 
particle physics community
some reasonably accurate numbers that were obtained 
directly from the QCD lagrangian,
with no other assumptions. But, it is wrong to present 
the two methods of including fermions,
one based on Kogut-Susskind fermions and the other 
based on lattice fermions with
exact chirality, as on equal theoretical footing. 
Nevertheless,~\cite{phystod} mentions unspecified
``cross checks'' between the two fermion methods, as if they were 
conceptually equivalent. The truth is that
once one can do simulations
with truly chiral fermions, all QCD work based on 
Kogut-Susskind fermions can be forgotten.

I disagree with the authors that LQCD has matured; 
rather, its practitioners have, and 
their relentless pursuit of computer resources seems to 
have drained some of them of 
the selfdiscipline required when presenting ones 
results to the rest of the particle
physics community. 
Statements like ``the most interesting lattice 
calculations with dynamical
simulations are the ones done by the MILC 
collaboration''~\cite{degrand}, written by
another member of MILC, are misleading even if later on 
in the paper one finds a technical discussion of caveats~\cite{degrand}.
In the same paper we read: ``Finally, because they are so big, 
lattice projects have a high profile. 
They cannot be allowed to fail.'' It is left unclear whether this speaks 
for the reliability of the results coming from these projects, or, against it.

Unlike experiments in nuclear or particle physics, lattice 
projects do not require large numbers
of people, so do not {\it have} to be big in terms of personnel. 
I think that only if it again becomes acceptable that some 
projects fail, and projects are not forced to be so big, 
does LFT stand a chance
to restore some of the status of reliability and respect 
it used to enjoy in theoretical physics. 
We should rethink the 
policy that concentrates almost all
of the computing power in the hands of few 
large collaborations. The past has
taught us that this has a tendency to stifle 
individual thinking and imaginative 
risk taking. People who broke the mold sometimes 
were penalized without scientific
justification and ended up leaving LFT. 
Now is a good time to rethink this 
policy because an alternative exists:
small but reasonably effective commodity clusters 
have reached prices that are reasonable
for small groups and even individual researchers. 
Money would be better spent if, say,
one half of the resources allocated to LFT were distributed 
to small research groups, or even single researchers, earmarked 
for purchasing compute-clusters. No science coming from the
large collaborations would be lost if their funds were restricted to
the remaining half of the entire computation budget of LFT.


\end{document}